\documentstyle[multicol,aps,prl,epsf]{revtex}

\begin{document}

\draft

\title{Conductance of a single molecule anchored by an isocyanide substituent to gold 
electrodes}

\author{Manabu Kiguchi, Shinichi Miura, Kenji Hara, Masaya Sawamura, Kei Murakoshi}

\address{Division of Chemistry, Graduate School of Science, Hokkaido University, Sapporo, 060-0810, 
Japan}

\date{\today}

\maketitle

\begin{abstract}

The effect of anchoring group on the electrical conductance of a single molecule 
bridging two Au electrodes was studied using di-substituted (isocyanide (CN-), thiol (S-) or 
cyanide (NC-)) benzene. The conductance of a single Au/1,4-diisocyanobenzene/Au junction 
anchored by isocyanide via a C atom (junction with the Au-CN bond) was 
$3 \times 10 ^{-3} G_{0}$ ($2e^{2}/h$). 
The value was comparable to  $4 \times 10 ^{-3} G_{0}$ of a single Au/1,4-benzenedithiol/Au junction with 
the Au-S bond. The Au/1,4-dicyanobenzene/Au molecular junction with the Au-NC bond did 
not show well-defined conductance values. The metal-molecule bond strength was estimated 
by the distance over which the molecular junction was stretched before breakdown. The 
stretched length of the molecular junction with the Au-CN bond was comparable to that of the 
Au junction, indicating that the Au-CN bond was stronger than the Au-Au bond. 
\end{abstract}

\medskip

\begin{multicols}{2}
\narrowtext

Recently, molecular electronic devices have attracted attention, since these devices offer the 
possibility for extreme scale down and ultra-small energy consumption of logic units. 
Building the device with a single molecule demands an understanding of the electron 
transport properties through single molecule junctions connecting two metal electrodes. The 
electron transport characteristics through a single molecule have been studied using 
mechanically controllable break junctions \cite{1} and scanning tunneling microscopy (STM) \cite{2,3,4}. 
With regard to anchoring groups of the molecular junction, the Au-S bond is used to connect 
molecules to metal electrodes. It is because the Au-S bond is a strong bond, which leads to the 
effective hybridization of the molecular and metal orbitals. The effective hybridization of the 
orbitals is essential for the high conductance of the molecular junction. On the other hand, the 
high conductance of the molecular junction also requires the condition of energy matching 
between the Fermi level of metal electrodes and conduction orbital of the molecule \cite{5}. A strong 
metal-molecule bond and a smaller energy difference are essential for the molecular junction 
to show high conductance. The energy of the molecular orbital varies with the end group of 
the molecule. It is important to develop new end groups other than a thiol to enable more 
stable and higher conductance single molecular junctions.

Among various end groups, isocyanideend groups are attractive. Isocyanides bind to 
transition metals via a strong coordination bond by effective donation to metal and back 
donation from metals. The strength of the metal-CN bond is comparable to that of the metal-S 
bond. Because of this effective interaction between the isocyanide group and metals, air stable 
isocyanide SAM films are formed on Au \cite{6}, as with the case of thiol SAMs \cite{7}. This strong 
metal-CN bond may contribute to increase the conductance of the junction. Recent theoretical 
calculations show promise for a molecular junction based on an isocyanide group \cite{8}. The 
conductance of the molecular junction binding a benzene molecule to Au atoms by isocyanide 
coordination (the molecular junction with the Au-CN bond) is close to that of the molecular 
junction with the Au-S bond. In the case of Ni electrodes, the conductance of molecular 
junctions with isocyanide end groups is higher than those of thiols \cite{8}. Despite these interests, 
the conductance of a single molecule with isocyanide end groups has not been studied 
adequately \cite{9}. 

In the present study, we have measured the electrical conductance of several 
di-substituted (-S, -NC, -CN) benzenes bridging two Au electrodes. As for the di-substituted 
benzene, there are only a few reports on the conductance of 1,4-benzenedithiol bridging two 
Au electrodes \cite{1,3,4}. Based on the bond strength of the anchoring group \cite{10}, 
the conductance of a single molecule was discussed using a simple tunneling model \cite{5}.

A sharp tip was used, comprising a Au wire (diameter $\sim$0.25 mm, $>$99.9$\%$). The substrate 
was Au(111), prepared by a flame annealing and quenching method \cite{11}. The measurements 
were carried out in tetraethyleneglycol dimethyl ether (tetraglyme) or toluene solutions since 
isocyanides could be hydrolyzed under aqueous acidic or basic conditions \cite{12}. The solution 
contained 1mM concentrations of 1,4-diisocyanobenzene, 1,4-benzenedithiol, or 
1,4-dicyanobenzene. A Au STM tip was repeatedly moved into and out of contact with a Au 
(111) substrate at a rate of 50 nm/s in the solution. Each molecule was terminated with two 
isocyanide, thiol, or cyanide end groups bridging the tip and substrate electrodes via the 
respective anchoring groups, as shown in Fig. 1. Conductance was measured during the 
breaking process under an applied bias of 20 mV between the tip and substrate. All statistical 
data was obtained from a large number (over 1000) of individual conductance traces.

Figure 2 (a,b) show the typical conductance traces of Au point contacts broken in the  
tetraglyme solution containing 1 mM of 1,4-diisocyanobenzene. The conductance decreased 
in a stepwise fashion with each step occurring at an integer multiple of 
$2 \times 10 ^{-3} G_{0}$ $\sim$ $3 \times 10 ^{-3} G_{0}$. 
The last plateau before the contact break showed a value of $2 \times 10 ^{-3} G_{0}$. 
The corresponding 
histogram shows a feature at  $3 \times 10 ^{-3} G_{0}$($\pm 1 \times 10 ^{-3} G_{0}$ ). 
These steps in the conductance trace 
likely originate from the formation of a stable, single Au/1,4-diisocyanobenzene/Au junction, 
possibly binding a benzene molecule to Au atoms by isocyanide coordination via C atoms 
(molecular junction with the Au-CN bond). The conductance behavior in the absence of 
1,4-diisocyanobenzene molecules was studied. Neither steps nor peaks were observed in the 
same conductance regime (see Fig. 2 (a,b)). The effect of the solvent was also checked using 
toluene. The inset of Fig. 2 (b) shows the conductance histogram observed in toluene solution 
containing 1 mM of 1,4-diisocyanobenzene. The conductance histogram shows a feature near 
$3 \times 10 ^{-3} G_{0}$, as is the case with tetraglyme. Based on these experimental results, the 
conductance of a single molecular junction with the Au-CN bond was determined to be 
$3 \times 10 ^{-3} G_{0}$($\pm 1 \times 10 ^{-3} G_{0}$ ).

The conductance of 1,4-dicyanobenzene and 1,4-benzenedithiol is discussed. In the 
Au/1,4-dicyanobenzene/Au junction (molecular junction with the Au-NC bond), a molecule 
binds to Au atoms via the N atom of the cyanide group, while a molecule binds to Au atoms 
via the S atom in the Au/1,4-benzenedithiol/Au molecular junction (molecular junction with 
the Au-S bond). Figure 2 (c-f) show the conductance traces and histograms of Au point 
contacts broken in the tetraglyme solution containing 1mM of 1,4-dicyanobenzene or 
1,4-benzenedithiol. As for 1,4-dicyanobenzene, small conductance steps were observed in the 
conductance trace. The difference in conductance between the steps was smaller than 
$5 \times 10 ^{-4} G_{0}$. The conductance histogram does not show a clear feature above 
$1 \times 10 ^{-3} G_{0}$. Thus, 
the conductance of the molecular junction with the Au-NC bond would be smaller than 
$1 \times 10 ^{-3} G_{0}$. Although another possibility that the molecule could not make stable and 
reproducible junctions should be also considered for the interpretation on this conductance 
behavior, estimated value may reflect relatively small conductivity of the junction. For 
1,4-benzenedithiol, steps occurring at an integer multiple of $4 \times 10 ^{-3} G_{0}$ 
were observed in the conductance traces, and the conductance histogram shows a feature 
near $4 \times 10 ^{-3} G_{0}$. 
Thus, the conductance of a single molecular junction with the Au-S bond was determined to be   
$4 \times 10 ^{-3} G_{0}$($\pm 1 \times 10 ^{-3} G_{0}$ ). 
The conductance of a single molecular junction with the Au-S bond was also 
studied by other groups, and it ranges from  $1.1 \times 10 ^{-2} G_{0}$ \cite{13} 
to $4.1 \times 10 ^{-4} G_{0}$ \cite{1}. Our result of 
$4 \times 10 ^{-3} G_{0}$ falls within this range of these values. 
Since the different conductance values for 
the same molecular junction by orders of magnitude may be due to the different experimental 
conditions, it is difficult to compare the conductance of single molecular junctions of different 
molecules measured under different conditions. In the present case, the conductance of a 
single molecular junction is discussed based on our results measured under the same 
condition. Statistical analysis of the present data showed that the conductance values of a 
single molecular junction measured under the same condition could be determined to be in a 
precision of 30 $\%$. Our results showed that the conductance of the molecular junction with the 
Au-CN bond was comparable to the junction with the Au-S bond. The conductance of the 
molecular junction with the Au-NC bond was much smaller than the junction with the Au-S 
and Au-CN bonds.

Our experimental results on the conductance of a single molecular junction is discussed 
with a simple tunneling model and theoretical calculation results. In the simple tunneling 
model \cite{5}, the conductance of a single molecular junction is given 
by $A(\beta G \rho)^{2}$, where $A$, $\beta$, $G$, $\rho$
are the constant, hopping integral between the molecule and metal, Green's function, and 
local density of states (LDOS) at the Fermi energy of the contact metal atom, respectively.   
$\beta$ increases with the bond strength. $G$ decreases with the energy difference between the Fermi 
level and conduction orbital. Therefore, a strong metal-molecule bond (large $\beta$) and small 
energy difference (large $G$) are essential for a molecular junction with high conductance. In 
the case of the molecular junction with the Au-CN bond, the bond strength and the energy 
difference between the Fermi level and conduction orbital are similar to that with the Au-S 
bond (similar $\beta$ and $G$) \cite{14,15}. Thus, the conductance of the molecular junction with the 
Au-CN bond would be close to that of the Au-S bond. Theoretical calculations show that the 
molecular junction with the Au-S bond and Au-CN bond have similar conductance values \cite{8}. 
Our experimental results agree well with those theoretical estimates. On the other hand, the 
strength of the Au-NC bond is much smaller than that of the Au-CN or Au-S bonds, which 
yields a small for the Au-NC bond. Since the energy difference between the Fermi level and 
conduction orbital of 1,4-dicyanobenzene is close to that of 1,4-diisocyanobenzene \cite{13,14}, the 
conductance of the molecular junction with the Au-NC bond would be much smaller than that 
with the Au-CN bond due to a small $\beta$. Theoretical calculations shows that the conductance 
of the molecular junction with the metal-NC bond is three times smaller than that with the 
metal-S bond \cite{14,15}. 

Relatively strong binding between Au and C atoms in Au-CN binding was also 
demonstrated by conductance traces. Figure 3 (a,b) show the typical conductance traces of Au 
point contacts broken in the presence of 1,4-diisocyanobenzene. The length of the last 
conductance plateau of the molecular junction (0.16 nm in Fig. 3 (b)) was comparable to that 
of the Au junction (0.12 nm in Fig. 3 (a)). Since the last plateau corresponds to the single 
molecular or metal atom junction \cite{11}, the length of the last plateau is the distance over which a 
molecular or metal junction can be stretched before breakdown (breakdown distance). To 
evaluate characteristics of the breakdown distance quantitatively, the distributions of the 
breakdown distance of the Au junctions and molecular junction are shown in Fig. 3 (c). The 
breakdown distance was defined as the distance between the points at which the conductance 
dropped below 1.2 $G_{0}$ and 0.8 $G_{0}$ for the Au junctions, and  
$3.2 \times 10 ^{-3} G_{0}$ and  $2.0 \times 10 ^{-3} G_{0}$ 
for the molecular junctions, respectively. The average length for the Au junctions and the 
molecular junction was 0.078 ($\pm$0.008) nm and 0.073 ($\pm$0.007) nm, respectively. Two 
breakdown distances agreed well with each other. 

The breakdown can take place at one of the three bonds, Au-CN, Au-Au, or an 
intramolecular bond in the Au/1,4-diisocyanobenzene/Au molecular junction. The respective 
bonds could show different breakdown distances because of the difference in the bond 
strength. Comparable breakdown distances indicates that breakdown in the molecular junction 
took place at the Au-Au bond rather than the Au-CN bond, which is compelling proof that the 
Au-CN bond was stronger than the Au-Au metal bond in the molecular junction with the 
Au-CN bond. In the case of the Au/1,4-benzenedithiol/Au molecular junction, the breakdown 
distance (0.10 nm in Fig. 2(c)) was also comparable to that of the Au junction, indicating that 
the Au-S bond was also stronger than the Au-Au bond. The relatively short length of the 
plateau (0.02 nm see Fig. 2 (e)) of the Au/1,4-dicyanobenzene/Au junction indicates that the 
Au-NC bond was weaker than the Au-Au bond. Similar behavior was also observed in the 
system of 1,8-octanedithiol bonded to two Au electrodes by force measurement using atomic 
force microscopy (AFM) \cite{16}. The force required to break the molecular junctions was close to 
that of the Au junctions, indicating that the strength of the Au-S bond is higher than that of the 
Au-Au metal bond. 

A well-defined molecular junction with a strong Au-CN bond was fabricated in the 
present study. As for the di-substituted benzene, the conductance of the molecular junction 
with the Au-CN bond was comparable to that with the Au-S bond. The energy of the 
molecular orbital varies with the linker group and the end group of the molecule. In the case 
of 1,4-diisocyanobenzene, the linker and end group are defined as benzene ring (C$_{6}$H$_{4}$) and 
isocyanide (CN), respectively. In some linker groups, the energy difference between the Fermi 
level of the metal electrodes and conduction orbital of the molecule could be smaller for 
isocyanide end groups than thiols. Since the bond strength of the metal-CN bond is 
comparable to that of the metal-S bond, the conductance of the molecular junction with the 
metal-CN bond could be higher than that with the metal-S bond in some combinations of the 
metals and molecules. In developing an electronic device with a single molecule, it will be 
important to study electron transport through a single molecule with a high degree of 
functionality, such as in the photoswitching effect and single molecule magnetism. By using 
the best end group for the individual molecule, the performance of the device would be 
drastically improved. The isocyanide end group would be a suitable candidate.

In conclusion, we studied the conductance of a single 1,4-diisocyanobenzene, 
1,4-dicyanobenzene or 1,4-benzenedithiol molecule bridging two Au electrodes. The 
conductance of the molecular junction with the Au-CN bond was comparable to that with the 
Au-S bond. The plateau length analysis showed that the Au-CN coordination bond was 
stronger than Au-Au metal bond. The isocyanide end group is a promising alternative to the 
thiol. This work was partially supported by a Grant-in-Aid for Scientific Research A (No. 
16205026) and Grant-in-Aid for Scientific Research on Priority Areas (No. 17069001) from 
MEXT.

\begin{figure}
\begin{center}
\leavevmode\epsfxsize=70mm \epsfbox{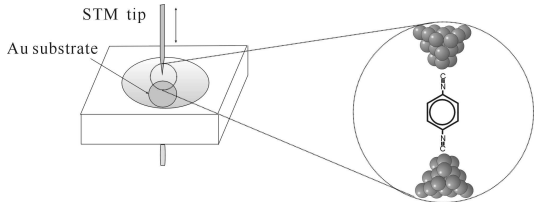}
\caption{
Schematic illustration of a molecule junction, in which a single molecule is bonded to 
two Au electrodes.}
\label{fig1}
\end{center}
\end{figure}

\begin{figure}
\begin{center}
\leavevmode\epsfxsize=85mm \epsfbox{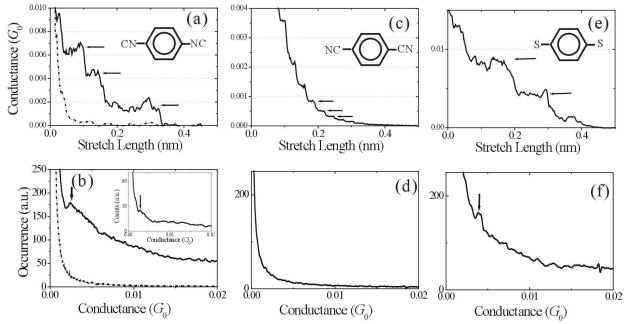}
\caption{
Typical conductance trace and histogram of Au point contacts broken in the tetraglyme 
solution consisting of (a,b) 1,4-diisocyanobenzene, (c,d) 1,4-dicyanobenzene and (e,f) 
1,4-benzenedithiol. The dotted-line shows the result in the absence of molecules. The inset of 
(b) shows the conductance histogram in the toluene solution containing 1 mM of 
1,4-diisocyanobenzene.}
\label{fig2}
\end{center}
\end{figure}

\begin{figure}
\begin{center}
\leavevmode\epsfxsize=70mm \epsfbox{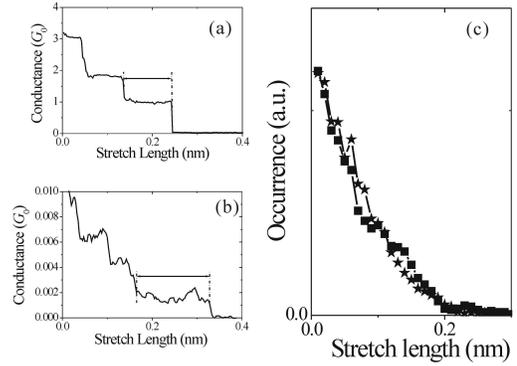}
\caption{Typical conductance trace of Au junctions (a) and 1,4-diisocyanobenzene junctions in 
tetraglyme solution containing 1 mM of 1,4-diisocyanobenzene. (c) The distribution of 
lengths for the last conductance plateau for the Au junctions (box), and molecular junctions 
(star).}
\label{fig3}
\end{center}
\end{figure}

\end{multicols}
\end{document}